\documentclass[prl,twocolumn,a4paper]{revtex4-1}
\usepackage{graphicx}
\usepackage{enumitem}
\usepackage{amsmath}
\newcommand{\ket}[1]{|{#1}\rangle}
\newcommand{\bra}[1]{\langle{#1}|}

\def\Xint#1{\mathchoice
   {\XXint\displaystyle\textstyle{#1}}%
   {\XXint\textstyle\scriptstyle{#1}}%
   {\XXint\scriptstyle\scriptscriptstyle{#1}}%
   {\XXint\scriptscriptstyle\scriptscriptstyle{#1}}%
   \!\int}
\def\XXint#1#2#3{{\setbox0=\hbox{$#1{#2#3}{\int}$}
     \vcenter{\hbox{$#2#3$}}\kern-.5\wd0}}

\def\dashint{\Xint-}
\usepackage{amsfonts}

\begin{document}

\title{Universal duality transformations in interacting one-dimensional quantum systems}

\author{Manuel Valiente}
\affiliation{Institute for Advanced Study, Tsinghua University, Beijing 100084, China}

\begin{abstract}
One-dimensional quantum systems admit duality relations that put hard core spinless bosons and fermions in one-to-one correspondence via Girardeau's mapping theorem. The simplest models of soft bosons interacting via zero-range potentials can also be mapped onto dual interacting fermions. However, a systematic approach to one-dimensional statistical transmutation for arbitrary low-energy interactions in the spinless and spinful or multicomponent cases has remained elusive. I develop a general theory of local unitary transformations between one-dimensional quantum systems of bosons and fermions with arbitrary spin or internal structure, single-particle dispersion -- including non-relativistic, relativistic or otherwise -- and low-energy interactions in the universal regime. These transformations generate families of new duality relations and models that relate the strong and weak coupling limits of the respective dual theories.                   
\end{abstract}
\pacs{}

\maketitle

Quantum wires, or many-body systems of quantum particles kinematically constrained to effectively move in one spatial dimension (1D), constitute prominent examples where interactions and quantum fluctuations are highly enhanced when compared to their three-dimensional counterparts \cite{Giamarchi2004}. Examples of experimental realizations of (quasi-) 1D quantum systems include trapped few- and many-body ensembles of ultracold atoms \cite{Jochim2015,Reynolds2020,Hulet2020,Bloch2004,Haller2009}, cold atomic $^4\mathrm{He}$ in nanopores \cite{Duc2015}, one-dimensional nanowires \cite{Auslaender2002} or the edge states of topological insulators \cite{Molenkamp2020}. A peculiarity of 1D systems is the similarity in many physical properties of bosons and fermions. In 1D, spinless bosons and fermions are dual to one another via Girardeau's mapping \cite{Girardeau1960} provided their two-body interactions feature a hard short distance core. Girardeau's mapping has been generalized to some multicomponent systems with zero-range hard cores \cite{Minguzzi2007,Deuretzbacher2008}, while fermionic duals of the integrable Lieb-Liniger model for spinless \cite{Cheon1999} and two-component bosons with lowest order interactions \cite{Girardeau2004} have also been found. All of these relations rely on the knowledge of microscopic solutions and short distance details, specific to the particular problem of interest. In this work, I develop a general theory of microscopic duality transformations between bosonic and fermionic systems in 1D. These apply regardless of internal structure or spin, single-particle dispersion, two- and higher-body interactions and their integrable or non-integrable nature.

Duality relations, when these exist, between systems \textit{A} and \textit{B} typically map the strong-coupling limit of \textit{A} into the weak-coupling regime of \textit{B} and viceversa \cite{Girardeau1960,Cheon1999,Kramers1941}. The only known form of Bose-Fermi duality in 1D without hard core interactions corresponds to lowest order two-body even- and odd-wave interactions \cite{Cheon1999,Girardeau2004,Sekino2018,Granger2005,Minguzzi2006} in effective field theory (EFT) \cite{Valiente2015,Braaten2008}, which describes interacting quantum systems in the so-called universal regime, where short-range details of the interactions are unimportant \cite{Kaplan2005}. The derivation of 1D duality relations has so far relied heavily on knowledge of non-interacting states and Hamiltonians, together with short-range boundary conditions. The latter are ultimately linked to the specific form of the Hamiltonian if the interactions come from a low-energy EFT. This approach is very complicated beyond the simplest of models. Here, I instead use operator methods that do not rely on microscopic details to derive duality relations between general systems of 1D bosons and fermions.

I introduce the concept of statistical transmutation operators (STO), denoted by $\mathcal{T}$. These formally transform functions with bosonic symmetry into functions with fermionic symmetry and viceversa. Consider a one-dimensional $N$-body system with arbitrary internal or spin structure, and denote the position coordinates of the particles by $x_i$ ($i=1,\ldots,N$). I require the STO to be (i) linear; (ii) unitary; (iii) local; and (iv) energy independent. Locality, i.e. $\bra{\mathbf{x}'}\mathcal{T}\ket{\mathbf{x}}\propto \delta(\mathbf{x}-\mathbf{x}')$, is required because if non-local unitaries are allowed then it is possible to formally choose one that diagonalizes the Hamiltonian. Energy independence of the STO is needed if the microscopic details of the Hamiltonian are {\it a priori} unknown.  

Bosonic and fermionic states shall be denoted, respectively, by $\ket{\psi}$ and $\ket{\chi}$ hereinafter. With the conditions (i)--(iv) above, the relations $\ket{\psi}=\mathcal{T}\ket{\chi}$ and $\ket{\chi}=\mathcal{T}^{\dagger}\ket{\psi}=\ket{\chi}$ hold. If the bosonic Hamiltonian is given by $H_{\mathrm{B}}$, then its fermionic dual is simply $H_{\mathrm{F}}=\mathcal{T}H_{\mathrm{B}}\mathcal{T}^{\dagger}$ or, more concretely, its projection onto the totally antisymmetric sector. If the non-interacting part of $H_{\mathrm{B}}$ is given by $H_0$, such that $H_{\mathrm{B}}=H_0+V_{\mathrm{B}}$, it is most useful and intuitive to rewrite its fermionic dual as $H^{(0)}_{\mathrm{F}}=H_0+W_{\mathrm{F}}$, where $W_{\mathrm{F}}$ is the totally antisymmetric projection of $W$, given by
\begin{equation}
  W=[\mathcal{T},H_0]\mathcal{T}^{\dagger}, \label{W}
\end{equation}
which will be referred to as the statistical interaction.

Consider first a general system of $N$ spinless bosons. The only candidate $\mathcal{T}$ for STO is
\begin{equation}
  \bra{\mathbf{x}'}\mathcal{T}\ket{\mathbf{x}} = S_N(\mathbf{x})\delta(\mathbf{x}-\mathbf{x}'),\label{STO1}\\
\end{equation}
where $S_N(\mathbf{x})=\prod_{i<j=1}^NS(x_{ij})$, with $x_{ij}=x_i-x_j$ and $S(x)$ the signum function. For $x_i\ne x_j$ $\forall i\ne j$, it holds that $S_N(\mathbf{x})^2=1$. However, the signum function $S(x)$, understood as a distribution \cite{Schwartz1951}, is undefined at $x=0$. Defining its square involves the product of distributions, generally undefined as generalized functions \cite{Schwartz1954}. This requires the introduction of an associative algebra of generalized functions, also necessary to unambiguously define products involving Dirac delta functions. Such an algebra was developed by Shirokov \cite{Shirokov1976}, denoted by $\mathcal{U}$. The product of distributions belonging to $\mathcal{U}$ has the defining properties  
\begin{align}
  [S(x)]^2&=1 \hspace{0.1cm} \forall x,\label{SignumSquared}\\
  [\delta(x)]^2&=0,\label{deltasquared}\\
  \{S(x),\delta(x)\}&=0.\label{anti}
\end{align}
Property (\ref{SignumSquared}) implies that the singular operator $\mathcal{T}$, Eq.~(\ref{STO1}), is unitary and therefore the STO for an arbitrary system of spinless bosons or fermions. 

In the non-relativistic spinless case, low-energy EFT of interactions, largely developed in a nuclear physics context \cite{Weinberg1990,Bedaque2002}, describes universal low-energy collisions in dimensionally reduced ultracold atomic physics in the two-body \cite{Valiente2015,Braaten2008,Bloch2008} and three-body sectors \cite{Hammer2013,Guijarro2018,Drut2018,Maki2019,Valiente2019,Nishida2018a,Nishida2018b,Pricoupenko2018}, as well as four-body \cite{Nishida2010}. For example, the ever present Dirac delta interaction in 1D is the lowest-order boson-boson interaction in EFT. I show now that bosonic and fermionic EFTs are, to each order, dual to each other.

I begin with the statistical interaction $W$, Eq.~(\ref{W}). In the position representation, it takes the form
\begin{equation}
  W(\mathbf{x})=-\frac{\hbar^2}{2m}S_N(\mathbf{x})\left[\nabla_N^2S_N(\mathbf{x})+2\nabla_NS_N(\mathbf{x})\cdot \nabla_N\right],
\end{equation}
where $\nabla_N$ is the $N$-dimensional gradient. Explicitly, one finds
\begin{equation}
  W(\mathbf{x})=-\frac{2\hbar^2}{m}\sum_{i<j=1}^NS(x_{ij})\left[\delta'(x_{ij})+2\delta(x_{ij})\partial_{x_{ij}}\right],
\end{equation} 
which is a pairwise interaction.

Two spinless bosons interact only through the even-wave part $V_{\mathrm{e}}^{(2)}$ of a general two-body interaction $V^{(2)}$. This is split into even- and odd-wave components as $V^{(2)}=V_{\mathrm{e}}^{(2)}+V_{\mathrm{o}}^{(2)}$. It is easy to see that
\begin{equation}
  \bra{x'_{ij}}V_{\mathrm{o}}^{(2)}\ket{x_{ij}}=S(x_{ij}')\bra{x'_{ij}}V_{\mathrm{e}}^{(2)}\ket{x_{ij}}S(x_{ij}).\label{VBtoVF}
\end{equation}
Relation (\ref{VBtoVF}) implies that the fermionic dual of the even-wave interaction is the odd-wave component of the original interaction $V^{(2)}$. In EFT, the expansion of the odd-wave interaction at low momentum transfer is given, in the momentum representation, by \cite{SOM}
\begin{equation}
  V_{\mathrm{o}}^{(2)}(k',k) = k'k\left[g_1+g_3(k^2+k'^2)+\ldots\right]\label{pwave}
\end{equation}
Its bosonic dual $V_{\mathrm{e}}^{(2)}=\mathcal{T}^{\dagger}V_{\mathrm{o}}^{(2)}\mathcal{T}$ is obtained in the momentum representation as
\begin{equation}
  V_{\mathrm{e}}^{(2)}(k',k)=\dashint \frac{\mathrm{d}q'}{\pi}\dashint \frac{\mathrm{d}q}{\pi} \frac{V_{\mathrm{o}}^{(2)}(q',q)}{(k'-q')(k-q)},\label{Hilbert}
\end{equation}
corresponding to a two-sided Hilbert transform \cite{King2009} ($\dashint$ is the principal value integral sign). To lowest order (LO), one has $g_{n\ge 3}\equiv 0$ in Eq.~(\ref{pwave}). If the bare LO coupling constant $g_1$ is kept finite, then the bosonic dual, Eq.~(\ref{Hilbert}), diverges. It is necessary to regularize the integral (\ref{Hilbert}), and I use hard cutoff ($\Lambda$) regularization. Expanding the integrals in powers of $\Lambda$ and $\Lambda^{-1}$, one finds that for LO bosonic interaction $V_{\mathrm{e}}^{(2)}(k',k)=g_0$, the bare coupling constant of its fermionic dual is given by $g_1=\pi^2g_0/4\Lambda^2$ as $\Lambda\to \infty$. In order to test this result, the odd-wave component of the statistical interaction in the momentum representation in the same regularization scheme is needed. One has, for two particles,
\begin{equation}
\bra{k'}W\ket{k}=\int\frac{\mathrm{d}q}{2\pi}S_{k'-q}^*\frac{\hbar^2q^2}{m}S_{k-q}-\frac{\hbar^2k^2}{m}\langle k' | k \rangle,\label{Wgen}
\end{equation}
where $S_k=\int \mathrm{d}x S(x)\exp(ikx)$ is the Fourier transform of the signum function. The odd-wave component $W_{\mathrm{o}}=W_{\mathrm{F}}$ is given by \cite{SOM}
\begin{equation}
W_{\mathrm{o}}(k',k)=-\frac{\pi\hbar^2}{m\Lambda}k'k,\label{WFkkp}
\end{equation}
which is also LO in EFT. Its even-wave component $W_{\mathrm{e}}(k',k)=4\hbar^2\Lambda/m\pi\to\infty$ would give the standard hardcore Bose-Fermi mapping in the absence of further interactions. The full LO fermion-fermion interaction $g_1k'k+W_F(k',k)\equiv g_Fk'k$ renormalizes and solves the bound state equation ($g_0<0$) with energy $E=-mg_0^2/4\hbar^2$ \cite{Lieb1963} in momentum space if, as $\Lambda\to \infty$, $1/g_F=-m\Lambda/\pi\hbar^2-(m/\hbar^2)^2g_0/4$. This value coincides with the inverse coupling constant derived from duality to $O(1)$. Importantly, renormalizability of the fermionic problem is guaranteed, by construction, from the duality transformation, since its bosonic dual is renormalizable.

Order by order, the fermionic dual interaction can be transformed into the bosonic EFT. To next-to-leading order (NLO), one has up to quadratic terms in the brackets of Eq.~(\ref{pwave}); for bosons one has $g_0+g_2(k^2+k'^2)$ instead. This is slightly more complicated than in LO. Denoting $a$ and $r$ the scattering length and effective range, respectively, one obtains dual bare coupling constants that depend non-trivially on the cutoff, $a$ and $r$ (See \cite{SOM}, where these are given explicitly). The procedure can be continued to any order.

The power of the duality transformation is even more patent with three-body interactions, which are of the same na{\"i}ve order as two-body effective range effects, and logarithmically modified upon renormalization \cite{Valiente2019}. At LO in the hyperradial momentum $q_{\mathrm{H}}$, the bosonic interaction is a constant in momentum space. For fermions, the LO interaction does not appear until the 6-th order. The LO bosonic problem in the absence of two-body interactions is renormalizable \cite{Guijarro2018,Drut2018,Maki2019,Valiente2019} with a single divergence in the ultraviolet (UV). For fermions, however, the interaction is highly singular and the Lippmann-Schwinger equation features three different divergent powers of a cutoff scale $\Lambda$ and a logarithmic divergence \cite{SOM}. This implies the irrelevance of three-body fermionic interactions in the absence of two-body forces. If two-body interactions are included, this is not necessarily the case, as is shown below by means of duality relations.

Consider the bosonic LO three-body interaction $V_{\mathrm{B}}^{(3)}(\mathbf{k}',\mathbf{k})=g_0^{(3)}$, with total momentum conservation implicitly assumed. The dual fermionic interaction is obtained as
\begin{equation}
  V_{\mathrm{F}}^{(3)}(\mathbf{k}',\mathbf{k})=g_0^{(3)}\int \frac{\mathrm{d}\mathbf{q}'}{(2\pi)^3}S^{\mathrm{A}*}_{\mathbf{q}',\mathbf{k}'}\int \frac{\mathrm{d}\mathbf{q}'}{(2\pi)^3}S^{\mathrm{A}}_{\mathbf{q},\mathbf{k}},\label{VF31}
\end{equation}
with total momentum conservation (a factor $2\pi\delta(K-K')$, $k=k_1+k_2+k_3$) implicitly assumed, $S^{\mathrm{A}}_{\mathbf{q},\mathbf{k}}=\mathcal{A}(S_{\mathbf{q}-\mathbf{k}}^{(3)})$, antisymmetrized with respect to $\mathbf{k}$, and $S_{\mathbf{k}}^{(3)}=\int \mathrm{d}\mathbf{x} S_3(\mathbf{x})\exp({i\mathbf{k}\cdot \mathbf{x}})$. The fermionic interaction (\ref{VF31}), using hard cutoff ($\tilde{\Lambda}$) regularization of the integrals, is given by \cite{SOM}
\begin{equation}
V_{\mathrm{F}}^{(3)}(\mathbf{k}',\mathbf{k})=\frac{g_0^{(3)}}{4\pi^2\tilde{\Lambda^6}}k_{12}'k_{13}'k_{23}'k_{12}k_{13}k_{23}+O(g_0^{(3)}\tilde{\Lambda}^{-8}),\label{VF32}
\end{equation}
which corresponds to a LO fermionic interaction. Since the statistical interaction $W_{\mathrm{F}}$ does not contain three-body terms, Eq.~(\ref{VF32}) is the total LO three-body fermionic interaction. Again, this procedure can be continued order by order, but higher order terms become quickly irrelevant. Three-body range, which may be important, can be included as an energy-dependent coupling constant \cite{Guijarro2018,Valiente2019}. This concludes the analysis of the spinless non-relativistic case, for which I have shown that bosons and fermions are equivalent to each other, order by order, in the universal regime of low-energy interactions.

%%%%%%%%%%%%%%%%%%%%%%%%%%%%%%%%%%%%%%%%%%%%%%%%%%%%%%%%%%
\begin{figure}[t]
\begin{center}
\includegraphics[width=0.49\textwidth]{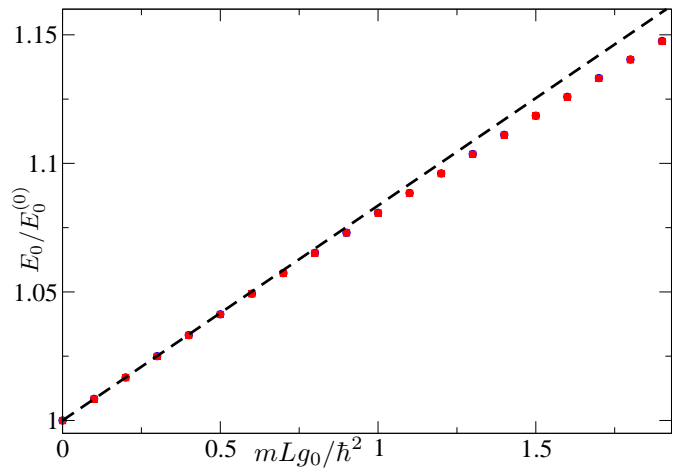}
\end{center}
\caption{Calculated ground state energy of three fermions (bosons) with spins $\uparrow\uparrow\downarrow$ (components $1$,$1$,$2$) corresponding to Yang's model and its dual, in a box of size $L$ with open boundary conditions as a function of the interaction strength $g_0$. Blue circles and red squares correspond, respectively, to fermions and bosons. Error bars are smaller than symbol sizes. Black dashed line is the weak-coupling result to $O(g_0)$\cite{McGuire1965}.}
\label{fig:Yang}
\end{figure}
%%%%%%%%%%%%%%%%%%%%%%%%%%%%%%%%%%%%%%%%%%%%%%%%%%%%%%%%%
Duality relations between multicomponent (spinful) bosons and spinful (multicomponent) fermions are obtained now. Because of the non-trivial matrix structure of the problem, more than one STO can be found. However, these are all unitarily equivalent to each other and I focus on the simplest one. Denote by $\mathbf{\xi}_i=(x_i,\mathbf{m}_i)$ ($i=1,\ldots,N$) the degrees of freedom of particle $i$, with $x_i$ the position and $\mathbf{m}_i$ a vector containing the internal degrees of freedom. An $N$-body wave function $\ket{\psi}$ satisfying bosonic statistics can be unitarily transformed into a wave function $\ket{\chi}$ satisfying fermionic statistics by means of the following local STO $\mathcal{T}$
\begin{align}
  \bra{\mathbf{\xi}_1',\ldots,\mathbf{\xi}_N'}\mathcal{T}\ket{\mathbf{\xi}_1,\ldots,\mathbf{\xi}_N}&=\delta(\mathbf{x}-\mathbf{x}')S_N(\mathbf{x})\prod_{i=1}^N\delta_{\mathbf{m}_i,\mathbf{m}_i'}.\label{diagonalSTO}
\end{align}
The duality relations for even- and odd-wave interactions, Eqs.~(\ref{Hilbert},\ref{VF31}), remain valid. However, they do not correspond to bosons and fermions as these are channel-dependent. For instance, spin-$1/2$ fermions are affected by odd-wave interactions only in the triplet channels, while the singlet channel is affected by the even-wave interaction. The statistical interaction $W$, on the other hand, depends heavily on the differential structure of the kinetic energy operator. For non-relativistic dispersion, the components of $W$ are obtained as in Eq.~(\ref{Wgen}), but is different otherwise. As an illustrative example, in Fig.~\ref{fig:Yang}, I show the numerically calculated \cite{SOM} ground state energy $E_0$ of three fermions with dynamics governed by Yang's Hamiltonian \cite{Yang1967,SOM} (dual bosons) with spin $\uparrow\uparrow\downarrow$ (components $1$,$1$,$2$) in a box of size $L$ with open boundary conditions in units of the fermionic non-interacting ground state energy $E_0^{(0)}$ as a function of the LO coupling constant $g_0$, where excellent agreement is clearly observed.

The duality relations are not restricted to the usual non-relativistic systems, but apply to general continuum Hamiltonians. Consider the following single-particle Hamiltonian, with two components $1$ and $2$, corresponding to the continuum limit of the Su-Schrieffer-Heeger (SSH) model \cite{Su1979} for spinless fermions near half-filling, 
\begin{equation}
  H_0=
\begin{pmatrix}
 0 & -i\hbar v\partial_x+i\delta\\
 -i\hbar v \partial_x-i\delta & 0  \\
\end{pmatrix}
.
\end{equation}
The components are identified with the two sublattices. In contrast with the usual non-relativistic cases considered so far, the Hamiltonian above is not channel-diagonal, contains only first derivatives, and its spectrum $E_{\lambda}(k)=(-1)^{\lambda}\sqrt{(\hbar v k)^2+\delta^2}$ ($\lambda=1,2$) is unbounded from below. For fermions, low-energy many-body physics is easy to handle as one can fill the infinite Fermi sea. For bosons, on the other hand, strong interactions are required beyond the few-body problem to obtain meaningful results. Therefore, a dual mapping to a fermionic problem is highly desirable in this and similar problems. The statistical interaction $W$ takes the simple form
\begin{equation}
  W(\mathbf{x})=-2i\sum_{i<j}S(x_{ij})\delta(x_{ij})\mathcal{M}_{ij},\label{WSSH}
\end{equation}
where $\mathcal{M}_{ij}$ is a Hermitian matrix acting on the sublattice degrees of freedom for the pair $(i,j)$,
\begin{equation}
\mathcal{M}_{ij}=\ket{11}(\bra{21}-\bra{12})+\ket{22}(\bra{12}-\bra{21})+\mathrm{H.c.},
\end{equation}
where $\ket{n_1n_2}\equiv \ket{n_1}_i\otimes\ket{n_2}_j$. In momentum space, $W$ is given by \cite{SOM}
\begin{equation}
W(k',k)=-\frac{4\pi\hbar v}{\Lambda}k\mathcal{M}_{ij}.
\end{equation}
Note that the momentum space structure of $W$ is different from the usual non-relativistic case, and it differs from the usual notion of hard core bosons, which would correspond to $W=g_0(\ket{11}\bra{11}+\ket{22}\bra{22})$, $g_0\to \infty$. That interaction would arise in the na\"ive continuum limit of the SSH model for strongly interacting bosons, which can be realized experimentally with ultracold atoms \cite{Atala2013}. The matrix $\mathcal{M}_{ij}$ mixes components in a non-trivial way. Since the coupling constant is strong, working with bosons in the many-body case is a hopeless venture. The fermionic dual, with no further interactions, corresponds to free fermions and is exactly solvable. Introducing LO even-wave interactions between fermions, of the form $g_0(\ket{12}\bra{12}+\ket{21}\bra{21})$, the bosonic dual is given by the usual odd-wave interaction $(\pi^2mg_0/4\hbar^2\Lambda^2)k'k(\ket{12}\bra{12}+\ket{21}\bra{21})$. The fermionic scattering amplitude is finite for finite $g_0$, and the bosonic scattering amplitude is also finite, albeit more complicated to obtain \cite{SOM}. 

Note that both the massive and massless (Luttinger) Thirring models \cite{Thirring1958} are mapped onto strongly-interacting bosons using the STO. These bosons are not to be confused with the bosons in Luttinger liquid theory \cite{Haldane1981} in the massless case, or with the bosons in sine-Gordon theory in the massive case \cite{Coleman1975}. Nonetheless, both fermionic models being exactly solvable, the STO provides an exactly solvable relativistic theory of strongly-interacting two-component bosons.

I now show that the statistical transmutation theory can be written as a gauge theory, which connects the 1D duality here presented with the 2D duality relevant for the Quantum Hall effect \cite{Wilczek1982,Zhang1989}. I introduce a singular gauge interaction $a_j$ ($j=1,2,\ldots,N$), which must be Hermitian and antisymmetric with respect to particle exchange, required to preserve Galilean invariance \cite{Aglietti1996} when this is present. For the Bose-Fermi mapping, the gauge interaction $a_j(\mathbf{x})$ is given by
\begin{equation}
  a_j(\mathbf{x})=2i\hbar\sum_{\ell=1}^NS(x_{j\ell})\delta(x_{j\ell}),
\end{equation}
where the sum above runs over $\ell\ne j$. The use of the properties of Shirokov's algebra, Eqs.~(\ref{SignumSquared},\ref{deltasquared},\ref{anti}), is crucial, for it implies Hermiticity of $a_j$, allows to set $S^2=1$ and, importantly, renormalizes squared delta functions to zero, which appear in the non-relativistic case.

Finally, the theory here described can be applied to one-dimensional anyons \cite{Keilmann2010,Greschner2015,Bonkhoff2020} with statistical parameter $\phi$ ($\phi=\pi/2$ corresponds to the Bose-Fermi mapping). The spatial part of the local STO is given by $\mathcal{T}_{\phi}(\mathbf{x})=i\exp\left[-i\phi\sum_{j<\ell}S(x_{j\ell})\right]$, which maps bosons (fermions) into anyons with angle $\phi$ ($\phi+\pi/2$). The derivation of statistical and dual interactions proceeds in a manner that is analogous to the Bose-Fermi mapping \cite{SOM}.

In conclusion, I have presented a general theory of Bose-Fermi statistical transmutation in quantum one-dimensional systems. In the low-energy regime, this applies to arbitrary systems of identical particles. As a consequence, essentially every result ever obtained in one dimension for bosons (fermions) in continuous space with low-energy interactions can be translated to fermions (bosons) with their respective statistical and dual interactions. These results may be used to characterize the formation of strongly-coupled spinful fermionic droplets by referring to available, simple analyses for multicomponent bosons \cite{Petrov2016,Astrakharchik2018}, which imply, via duality, that spin-$1/2$ fermionic droplets and liquids exist for strongly attractive odd-wave interactions in the triplet channels, or to study coupled wires \cite{Kane2002} fully microscopically, and the continuum limits of optical lattice-based ladders \cite{Celi2014,Tai2017} for bosons and fermions in a unified manner.


\begin{thebibliography}{99}%
\bibitem{Giamarchi2004} T. Giamarchi, {\it Quantum physics in one dimension} (Oxford Univ. Press, 2004).
  
\bibitem{Jochim2015} S. Murmann, F. Deuretzbacher, G. Z{\"u}rn, J. Bjerlin, S.~M. Reimann, L. Santos, T. Lompe and S. Jochim, Phys. Rev. Lett. {\bf 115}, 215301 (2015).

\bibitem{Reynolds2020} L.~A. Reynolds, E. Schwartz, U. Ebling, M. Weyland, J. Brand and M.~F. Andersen, Phys. Rev. Lett. {\bf 124}, 073401 (2020).

\bibitem{Hulet2020} Y.~-T. Chang, R. Senaratne, D. Cavazos-Cavazos and R.~G. Hulet, e-print arXiv:2007.03723 (2020).

\bibitem{Bloch2004} B. Paredes, A. Widera, V. Murg, O. Mandel, S. F{\"o}lling, I. Cirac, G.~V. Shlyapnikov, T.~W. H{\"a}nsch and I. Bloch, Nature {\bf 429}, 277 (2004).

\bibitem{Haller2009} E. Haller, M. Gustavsson, M.~J. Mark, J.~G. Danzl, R. Hart, G. Pupillo and H.~C. N{\"a}gerl, Science {\bf 325}, 1224 (2009).

\bibitem{Duc2015} P.~F. Duc, M. Savard, M. Petrescu, B. Rosenow, A. Del Maestro and G. Gervais, Science Adv. {\bf 1}, e1400222 (2015).

\bibitem{Auslaender2002} O.~M. Auslaender, A. Yacoby, R. de Picciotto, K.~W. Baldwin, L.~N. Pfeiffer and K.~W. West, Science {\bf 295}, 825 (2002).

\bibitem{Molenkamp2020} J. Strunz, J. Wiedenmann, C. Fleckenstein, L. Lunczer, W. Beugeling, V.~L. M{\"u}ller, P. Shekhar, N.~T. Ziani, S. Shamim, J. Kleinlein, H. Buhmann, B. Trauzettel and L.~W. Molenkamp, Nature Phys. {\bf 16}, 83 (2020).

\bibitem{Girardeau1960} M. Girardeau, J. Math. Phys. {\bf 1}, 516 (1960).

\bibitem{Minguzzi2007} M.~D. Girardeau and A. Minguzzi, Phys. Rev. Lett. {\bf 99}, 230402 (2007).

\bibitem{Deuretzbacher2008} F. Deuretzbacher, K. Fredenhagen, D. Becker, K. Bongs, K. Sengstock and D. Pfannkuche, Phys. Rev. Lett. {\bf 100}, 160405 (2008).

\bibitem{Lieb1963} E.~H. Lieb and W. Liniger, Phys. Rev. {\bf 130}, 1605 (1963).
  
\bibitem{Cheon1999} T. Cheon and T. Shigehara, Phys. Rev. Lett. {\bf 82}, 2536 (1999).

%\bibitem{Haldane1981} F.~D.~M. Haldane, J. Phys. C: Solid State Phys. {\bf 14}, 2585 (1981).

%\bibitem{Imambekov2009} A. Imambekov and L.~I. Glazman, Science {\bf 323}, 228 (2009).

%\bibitem{Abrikosov1975} A.~A. Abrikosov, L.~P. Gorkov and I.~E. Dzyaloshinski, {\it Methods of quantum field theory in statistical physics} (Dover 1975).

\bibitem{Girardeau2004} M.~D. Girardeau and M. Olshanii, Phys. Rev. A {\bf 70}, 023608 (2004).

\bibitem{Sekino2018} Y. Sekino, S. Tan and Y. Nishida, Phys. Rev. A {\bf 97}, 013621 (2018).

\bibitem{Granger2005} S.~A. Bender, K.~D. Erker and B.~E. Granger, Phys. Rev. Lett. {\bf 95}, 230404 (2005).

\bibitem{Minguzzi2006} M.~D. Girardeau and A. Minguzzi Phys. Rev. Lett. {\bf 96}, 080404 (2006).
  
\bibitem{Kramers1941} H.~A. Kramers and G.~H. Wannier, Phys. Rev. {\bf 60}, 252 (1941).
  
\bibitem{Valiente2015} M. Valiente and N.~T. Zinner, Few-body Syst. {\bf 56}, 845 (2015).

\bibitem{Braaten2008} E. Braaten, M. Kusunoki and D. Zhang, Ann. Phys. (NY) {\bf 323}, 1770 (2008).
  
\bibitem{Kaplan2005} D. Kaplan, Lectures delivered at the 17th National Nuclear Physics Summer School 2005, Berkeley, CA, June 6-17, 2005; e-print arXiv:051023.
  
\bibitem{Schwartz1951} L. Schwartz, {\it Th\'eorie des distributions} (Hermann, 1951).

\bibitem{Schwartz1954} L. Schwartz, C. R. Acad. Sci. Paris {\bf 239}, 847 (1954).

\bibitem{Shirokov1976} Yu.~M. Shirokov, Teor. Mat. Ftz. {\bf 28}, 308 (1976).  

\bibitem{Weinberg1990} S. Weinberg, Phys. Lett. B {\bf 251}, 288 (1990).

\bibitem{Bedaque2002} P.~F. Bedaque and U. van Kolck, Ann. Rev. Nucl. Part. Sci. {\bf 52}, 339 (2002).

%\bibitem{Hammer2013} H.-~W. Hammer, A. Nogga and A. Schwenk, Rev. Mod. Phys. {\bf 85}, 197 (2013).

\bibitem{Bloch2008} I. Bloch, J. Dalibard and W. Zwerger, Rev. Mod. Phys. {\bf 80}, 885 (2008).

\bibitem{Hammer2013} H.-~W. Hammer, A. Nogga and A. Schwenk, Rev. Mod. Phys. {\bf 85}, 197 (2013).

\bibitem{Guijarro2018} G. Guijarro, A. Pricoupenko, G.~E. Astrakharchik, J. Boronat and D.~S. Petrov, Phys. Rev. A {\bf 97}, 061605 (2018).

\bibitem{Drut2018} J.~E. Drut, J.~R. McKenney, W.~S. Daza, C.~L. Lin and C.~R. Ord{\'o}{\~n}ez, Phys. Rev. Lett. {\bf 120}, 243002 (2018).

\bibitem{Maki2019} J. Maki and C.~R. Ord{\'o}{\~n}ez, Phys. Rev. A {\bf 100}, 063604 (2019).

\bibitem{Valiente2019} M. Valiente, Phys. Rev. A {\bf 100}, 013614 (2019).

\bibitem{Nishida2018a} Y. Sekino and Y. Nishida, Phys. Rev. A {\bf 97}, 011602 (2018).

\bibitem{Nishida2018b} Y. Nishida, Phys. Rev. A {\bf 97}, 061603 (2018).

\bibitem{Nishida2010} Y. Nishida, D.~T. Son, Phys. Rev. A {\bf 82}, 043606 (2010).

\bibitem{Pricoupenko2018} L. Pricoupenko, Phys. Rev. A {\bf 97}, 061604 (2018).
  
\bibitem{King2009} F.~W. King, {\it Hilbert Transforms, Vol. 1} (Cambridge Univ. Press, 2009).

\bibitem{Yang1967} C.~N. Yang, Phys. Rev. Lett. {\bf 19}, 1312 (1967).

\bibitem{McGuire1965} J.~B. McGuire, J. Math. Phys. {\bf 6}, 432 (1965).
  
\bibitem{Su1979} W.~P. Su, J.~R. Schrieffer and A.~J. Heeger, Phys. Rev. Lett. {\bf 42}, 1698 (1979).

\bibitem{Atala2013} M. Atala, M. Aidelsburger, J.~T. Barreiro, D. Abanin, T. Kitagawa, E. Demler and I. Bloch, Nature Phys. {\bf 9}, 795 (2013).

\bibitem{Thirring1958} W. Thirring, Ann. Phys. {\bf 3}, 91 (1958).

\bibitem{Haldane1981} F.~D.~M. Haldane, J. Phys. C: Solid State Phys. {\bf 14}, 2585 (1981).
  
\bibitem{Coleman1975} S. Coleman, Phys. Rev. D {\bf 11}, 2088 (1975).

\bibitem{Wilczek1982} F. Wilczek, Phys. Rev. Lett. {\bf 49}, 957 (1982).

\bibitem{Zhang1989} S.~C. Zhang, T.~H. Hansson and S. Kivelson, Phys. Rev. Lett. {\bf 62}, 82 (1989).
  
\bibitem{Rabello1996} S.~J.~B. Rabello, Phys. Rev. Lett. {\bf 76}, 4007 (1996).
  
\bibitem{Aglietti1996} U. Aglietti, L. Griguolo, R. Jackiw, S.-~Y. Pi and D. Seminara, Phys. Rev. Lett. {\bf 77}, 4406 (1996).

\bibitem{Keilmann2010} T. Keilmann, S. Lanzmich, I. McCulloch and M. Roncaglia, Nature Comms. {\bf 2}, 361 (2010).

\bibitem{Greschner2015} S. Greschner and L. Santos, Phys. Rev. Lett. {\bf 115}, 053002 (2015).

\bibitem{Bonkhoff2020} M. Bonkhoff, K. J{\"a}gering, S. Eggert, A. Pelster, M. Thorwart and T. Posske, e-print arXiv:2008.00003.
  
\bibitem{Petrov2016} D.~S. Petrov and G.~E. Astrakharchik, Phys. Rev. Lett. {\bf 117}, 100401 (2016).

\bibitem{Astrakharchik2018} G.~E. Astrakharchik and B.~A. Malomed, Phys. Rev. A {\bf 98}, 013631 (2018).

\bibitem{Kane2002} C.~L. Kane, R. Mukhopadhyay and T.~C. Lubensky, Phys. Rev. Lett. {\bf 88}, 036401 (2002).

\bibitem{Celi2014} A. Celi, P. Massignan, J. Ruseckas, N. Goldman, I.~B. Spielman, G. Juzeliunas and M. Lewenstein, Phys. Rev. Lett. {\bf 112}, 043001 (2014).

\bibitem{Tai2017} M.~E. Tai, A. Lukin, M. Rispoli, R. Schittko, T. Menke, D. Borgnia, P.~M. Preiss, F. Grusdt, A.~M. Kaufman and M. Greiner, Nature {\bf 546}, 519 (2017).
  
\bibitem{SOM} See accompanying article: M. Valiente, Phys. Rev. A (to appear); e-print arXiv:2009.00624 .

%\bibitem{ShuChen2010} L. Guan and S. Chen, Phys. Rev. Lett. {\bf 105}, 175301 (2010).
  



  
\end{thebibliography}
\end{document}